\documentclass[conference]{IEEEtran}
\IEEEoverridecommandlockouts
\usepackage{cite}
\usepackage{amsmath,amssymb,amsfonts}
\usepackage{graphicx}
\usepackage{xcolor}

\usepackage{adjustbox}
\usepackage{tabularray}
\usepackage{url}
\usepackage{multirow}

\def\BibTeX{{\rm B\kern-.05em{\sc i\kern-.025em b}\kern-.08em
    T\kern-.1667em\lower.7ex\hbox{E}\kern-.125emX}}

\makeatletter
\newcommand{\newlineauthors}{%
  \end{@IEEEauthorhalign}\hfill\mbox{}\par
  \mbox{}\hfill\begin{@IEEEauthorhalign}
}
\makeatother

\title{Self-supervised Multimodal Speech Representations for the Assessment of Schizophrenia Symptoms
}

\author{\IEEEauthorblockN{Gowtham Premananth and  Carol Espy-Wilson\thanks{This work was supported by the National Science Foundation grant numbered 2124270.}}
\IEEEauthorblockA{
\textit{Dept. of Electrical and Computer Engineering, University of Maryland College Park, MD, USA},
\\
gowtham8@umd.edu,
espy@umd.edu\vspace{-2em}}

}

\begin{document}
\maketitle
\vspace{-12pt}
\begin{abstract}
Multimodal schizophrenia assessment systems have gained traction over the last few years. This work introduces a schizophrenia assessment system to discern between prominent symptom classes of schizophrenia and predict an overall schizophrenia severity score. We develop a Vector Quantized Variational Auto-Encoder (VQ-VAE) based Multimodal Representation Learning (MRL) model to produce task-agnostic speech representations from vocal Tract Variables (TVs) and Facial Action Units (FAUs). These representations are then used in a Multi-Task Learning (MTL) based downstream prediction model to obtain class labels and an overall severity score. The proposed framework outperforms the previous works on the multi-class classification task across all evaluation metrics (Weighted F1 score, AUC-ROC score, and Weighted Accuracy). Additionally, it estimates the schizophrenia severity score, a task not addressed by earlier approaches.
\end{abstract}

\vspace*{-8pt}

\section{Introduction}
\vspace*{-4pt}
Speech-based biomarkers have been used to detect and assess various disorders and medical conditions like Major Depressive Disorder, Dementia, and Parkinson's Disease \cite{Vikram2022}. The way that speech is affected by these psychiatric and psychological conditions has been utilized by the latest state-of-the-art deep learning-based speech processing techniques to create decision support systems and assessment tools for such disorders. Schizophrenia is one such mental health disorder that exhibits a variety of symptoms ranging from delusions and hallucinations to diminished emotional expression and poverty of speech. Previous works have shown that because of this heterogeneity of the symptoms presented by schizophrenia subjects and how it affects speech makes speech a well-suited biomarker for the detection and assessment of schizophrenia \cite{yashish_schizo,premananth2024multimodal}. Additionally, as human perception of speech has been considered multimodal involving audio-visual cues, these multimodal cues of speech have been used as biomarkers in various mental health assessment systems \cite{Nadee,HE202256}.

Clinicians use a variety of questionnaires and assessment techniques to measure the wide range of symptoms commonly exhibited by subjects with mental health disorders. The Brief Psychiatric Rating Scale (BPRS) \cite{overall1962brief} is one such questionnaire-based assessment tool used by clinicians to measure symptoms associated with schizophrenia. The BPRS scale consists of 18 such symptoms which are scored in a range of 1-7 depending on the severity of the symptom exhibited. The larger the score, the more severe is the symptom. These symptoms are mainly sub-typed into 3 prominent symptom classes strong Positive, strong Negative, and Mixed symptoms \cite{10.1001/archpsyc.1982.04290070025006}.

Previous studies on schizophrenia are limited to a binary classification or a multi-class classification problem. However, due to the diverse and varying nature of the symptoms in schizophrenia subjects, a real-world application like a clinical decision support tool needs to predict the severity of the symptoms. This needed expansion is one of the motivations for this work as severity prediction is not easily achievable given the limited access to medical data obtained from subjects.

Multimodal systems have been used in various domains with improved performance by the utilization of information extracted from multiple modalities. However, rather than extracting information from the modalities separately, multimodal representations have been developed in the recent past to acquire more information from modalities in a combined manner. Multimodal representation learning (MRL) is a learning technique to create multimodal representations that embed information from individual modalities that are not present in other modalities into a common representation \cite{Guo2019}. 

Multimodal speech representations have been mainly used for tasks like automatic speech recognition and audio-visual speaker diarization \cite{zhu2024multichannelavwav2vec2frameworklearning,shi2022learningaudiovisualspeechrepresentation}. Such multimodal speech representations have been mainly produced using self-supervised input features like Wav2Vec2.0\cite{baevski2020wav2vec20frameworkselfsupervised} and HuBERT\cite{hsu2021hubertselfsupervisedspeechrepresentation}. As these self-supervised representations lack interpretability, they cannot be used in systems developed with the goal of being deployed in the medical domain. For interpretability, we propose multimodal speech representations based on vocal Tract variables (TVs) \cite{Siriwardena2022,yashish_schizo} and Facial Action Units (FAUs) \cite{FACS}  that can be used for systems in the clinical domain. In addition, we show how these multimodal speech representations can effectively be used in the assessment of schizophrenia symptoms.

The \textbf{key contributions} of this work are as follows: 
\begin{enumerate}
\vspace*{-3pt}
    \item TV and FAU-based self-supervised multimodal speech representations that are task-agnostic.
    \item A Multi-Task Learning (MTL) based downstream model that uses the produced multimodal speech representations to estimate an overall schizophrenia severity score based on the BPRS scale and a classification output that classifies each input into a symptom-based class.
    \item A baseline for the speech-based schizophrenia severity score estimation.
\end{enumerate}

\vspace*{-8pt}
\begin{figure*}[h!]
    \centering
    \includegraphics[width=160mm, height= 50mm]{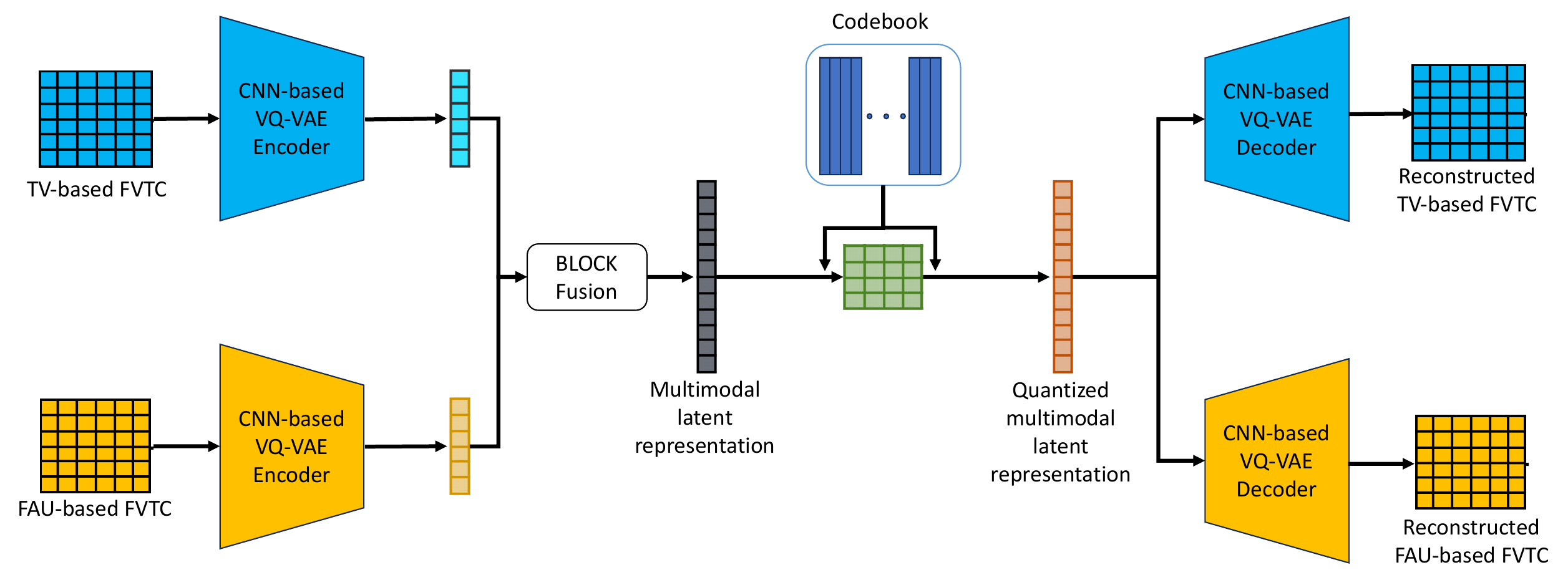}
    \caption{ MM-VQ-VAE based MRL model architecture}
    \label{fig:mrl}
\end{figure*}

\section{Dataset}
\vspace*{-4pt}
The multimodal dataset used in this study was collected at the University of Maryland School of Medicine in collaboration with the University of Maryland, College Park for a mental health study involving subjects with schizophrenia, depression, and healthy controls \cite{KELLY2020113496}. The dataset consists of video and audio recordings of subjects providing spontaneous responses in an unstructured interview with a clinical investigator. As this work focuses on schizophrenia, a subset of the multimodal dataset was used. The subset contains subjects exhibiting varying symptoms of schizophrenia and healthy controls who cover the whole spectrum of the BPRS scale. 

As this work focused on the two tasks mentioned above, the dataset needed two distinct targets: a severity score and a class label. The BPRS score provided for the session by the clinician before each interview session was considered the overall schizophrenia severity score. Also, the subjects were grouped into the symptom-based sub-types, Healthy Controls (HC), Positive Symptoms (P-SZ), and mixed symptoms (M-SZ). For example, the positive symptoms sub-type consists of the BPRS items grandiosity, hallucinatory behavior, conceptual disorganization, and unusual thought content \cite{shafer2005meta}. If the average severity scores for the items belonging to the symptom-based sub-types was above 3.5 then the subject was categorized into that sub-type. Note that the dataset didn't contain any subjects who exhibited strong Negative symptoms. The overall statistics of the dataset used for this study are provided in Table \ref{tab:dataset}. 

\vspace*{-8pt}
\begin{table}[h]
  \caption{\centering
  \textbf{Details of the Dataset used}}
  \label{tab:dataset}
  \centering
  \begin{tabular}{ |l | c | }
  \hline
    No.of Subjects&46\\
    \hline
    No.of Sessions&140\\
    \hline
    Hours of Speech&34.45\\
    \hline
    Symptom-based subclasses included & HC, P-SZ, M-SZ\\
    \hline
    BPRS-based severity score range &19 - 62\\
\hline
  \end{tabular}
\end{table}
\vspace*{-8pt}

\section{Data Pre-Processing and Feature Extraction}
\vspace*{-4pt}
The audio and video data in the dataset are of varying durations for each session and contain the speech of the interviewer and the subject. As the first step of pre-processing, the sessions were diarized based on the transcriptions of the sessions produced by a third-party transcription service, and the subjects' speech was separated. After that, all the sessions were segmented into 40-second segments using a 40-second window with a stride length of 5 seconds overlap. The segments were then used for both audio and video feature extraction.

From the segmented audio and video samples, TVs and FAUs were selected as low-level feature representations because they have been proven to give better performances when compared to other representations in similar multimodal mental health assessment systems \cite{yashish_schizo} and also contain some degree of interpretability.

The facial action coding system \cite{FACS} is a comprehensive anatomically based system for tracking facial movements which then converts individual muscle movement during facial movements into FAUs (video-based representation). Openface 2.0 toolkit \cite{Baltrusaitis2018} is used to extract FAUs from the videos. Out of all the FAUs extracted, 10 FAUs focused around the eyes, nose, and lip regions were selected to be used as input for the models. These FAUs yielded promising results in \cite{yashish_schizo,premananth24_interspeech} that used similar datasets and focused on mental-health identification tasks.

TVs are audio-based feature representations calculated using geometrical transformations of the location and degree of constriction formed by various articulators along the vocal tract. A total of 6 TVs are extracted using an acoustic-to-articulatory speech inversion system \cite{Siriwardena2022}.
Apart from the TVs, source information (Aperiodicity, Periodicity, and Pitch) was also extracted using the APP detector \cite{Deshmukh2005}. With all the 6 TVs extracted, 2 source features Aperiodicity and Periodicity were used as input for the models based on the performance they have shown in previous work \cite{yashish_schizo,premananth24_interspeech}.

\vspace*{-2pt}
\begin{table*}[th!]
\centering
   \caption{\textbf{Performance comparison of the proposed models with previous works}}
   \label{tab:performance}
    \begin{tabular}{|c | c| c | c | c | c|}
    \hline
    \multirow{2}{*}{\textbf{Modalities}} & \multirow{2}{*}{\textbf{Model}} & \multicolumn{3}{c|}{\textbf{Classification}} & \textbf{Regression} \\ \cline{3-6} 
                          &                                                    & \textbf{W.Acc$\uparrow$}     & \textbf{W.F1$\uparrow$}      & \textbf{AUC\_ROC$\uparrow$}      & \textbf{MAE$\downarrow$}       \\
\hline
Audio,Video&CNN-LSTM with Cross-Attention \cite{premananth2024multimodal}&52.17&49.33&68.53&-\\
    \hline
Audio,Video,Text&CNN-LSTM with Cross-Attention \cite{premananth2024multimodal}&56.52&51.62&74.21&-\\
    \hline
Audio,Video&CNN-LSTM with mGMU \cite{premananth24_interspeech}&60.87&60.28&77.35&-\\
\hline
Audio,Video,Text&CNN-LSTM with mGMU \cite{premananth24_interspeech}&65.22&65.47&82.14&-\\
\hline
Audio,Video&MM-VQ-VAE representations based classification model&54.96&71.04&57.62&-\\
\hline
Audio,Video&MM-VQ-VAE representations based regression model&-&-&-&8.81\\
\hline
    \textbf{Audio,Video}&\textbf{MM-VQ-VAE representations based  MTL model}&\textbf{75.00}&\textbf{76.41}&\textbf{91.52}&\textbf{7.19}\\
\hline
\end{tabular} 
\end{table*}

High-level features called Full Vocal Tract Coordination (FVTC) features were extracted from the selected TVs and FAUs separately (TV-based FVTC and FAU-based FVTC) based on the work by Huang et al \cite{Huang}. The FVTCs are correlation matrix structures based on the channel-delay correlation of the features. These FVTCs provide a more concise and well-informed representation of the TVs and FAUs based on their changes over time with respect to themselves and other features as well. These TV-based FVTC and FAU-based FVTC structures were used as input for the multimodal speech representation learning model.

\section{Methodology}
\vspace*{-4pt}
The methodology of this work can be explained in two stages.
\begin{enumerate}
  \item Self-supervised multimodal (audio-visual) speech representation learning to create task-agnostic representations.
  \item Multi-Task Learning (MTL) based downstream model to obtain 3-class classification labels and a severity score prediction using stacked multimodal speech representations.
\end{enumerate}

\subsection{Self-supervised Multimodal Speech Representation Learning (MRL)}
 \vspace*{-4pt}
Out of all the different types of MRL techniques \cite{Guo2019} such as joint representation, coordinated learning, and auto-encoder frameworks, the self-supervised auto-encoder framework approach was chosen for this work. We chose this MRL technique because this framework does not need labeled data for training which will allow the usage of data from multiple datasets. In future work, we plan to expand this work towards a more generalizable representation learning framework that can be used for various downstream tasks.

Generally, Variational Auto-Encoders (VAEs) consist of three distinct parts, an Encoder network that parameterizes a posterior distribution of $q(z|x)$ using discrete latent random variables $z$ given the input data $x$, a prior distribution $p(z)$, and a decoder with distribution $p(x|z)$. But in this work, instead of using conventional VAEs which produce a probabilistic latent space, the VQ-VAEs \cite{Oord2017} that produce a discrete latent space were used. The VQ-VAEs consider the posterior and prior distributions to be categorical and the samples are drawn from these distributions index an embedding table. These quantized embeddings are then used as input into the decoder network. The main reasoning behind the use of VQ-VAEs over conventional VAEs is that the discrete latent space produced in the VQ-VAEs will allow for more interpretable latent representations.

The VQ-VAE-based MRL architecture is shown in Fig.\ref{fig:mrl}. The CNN-based encoders used in this architecture comprise 2 2d-strided convolution layers, 2 2d-residual convolution layers, a 2d-convolution projection layer which are followed by a flattening layer, and a linear projection layer to create the latent representations. Then the unimodal latent representations obtained from the encoders are fused using BLOCK Fusion (Bilinear Superdiagonal Fusion)\cite{benyounes2019blockbilinearsuperdiagonalfusion}. Then the fused multimodal latent representations are quantized using the VQ-VAE codebook and then the quantized representations are used as input to both audio and video decoders. The decoders have the same set of layers arranged in the inverse order to reconstruct the input from the latent representation.

\vspace*{-8pt}
\begin{figure}[h!]
    \includegraphics[width=90mm, height= 25mm]{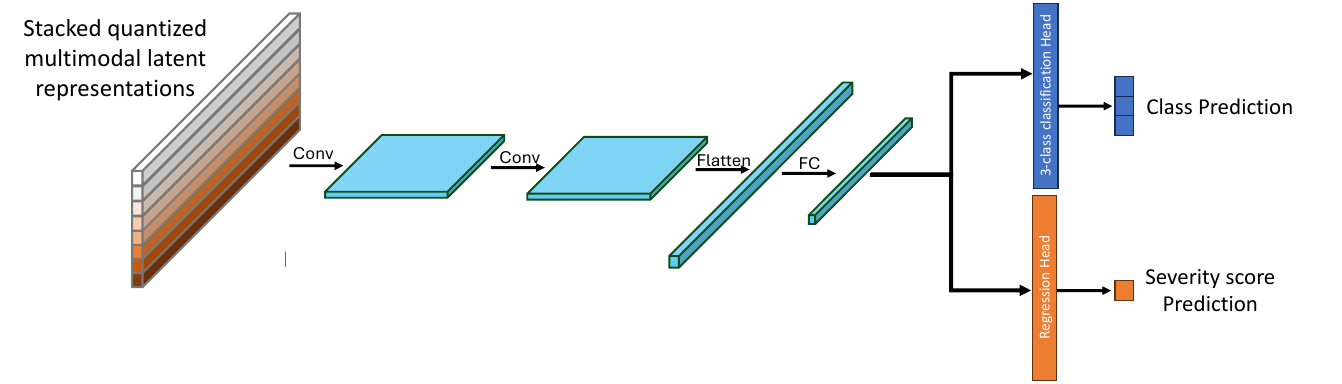}
    \caption{MM-VQ-VAE representation based MTL downstream model}
    \label{fig:downstream} 
\end{figure}

\subsection{Multi-task learning paradigm-based downstream model}
\vspace*{-4pt}

MTL \cite{MTL} is a learning paradigm that is often used in machine learning, where a model is trained not to do a single task but to perform multiple related tasks where the learning aspect of the model is enhanced due to the relatable nature of the tasks. This approach generally improves the model performance across the tasks and helps in deploying a single model for all the tasks with just multiple task-specific heads at the end rather than training and deploying separate models for each task. The important aspect of training a model with a MTL paradigm is the weighting of loss functions of the tasks during training. Even though a lot of different approaches have been used, we chose to use Homoscedastic Uncertainty Dynamic Loss Weighting \cite{Kendall2017MultitaskLU} where weights to the loss functions are applied based on the uncertainty of the tasks. In our case, we have a classification task and a regression task. The losses for the tasks were weighted by two learnable parameters in the model $\sigma_1$ and $\sigma_2$ that represented the uncertainty for each task during the training. The total loss function is computed using the following equation:
\\
\\
$L_{total}={\frac{L_{Classification}}{2(\sigma_1)^2}}+log(\sigma_1)+\frac{L_{Regression}}{2(\sigma_2)^2}+log(\sigma_2)$

\section{Experiments}
\vspace*{-4pt}

For training, testing, and validation purposes of all the models the dataset was divided into 70:15:15 train, test, and validation splits in a subject-independent manner with all the sessions belonging to a subject kept within a single split. The K-fold cross-validation approach that provides a more reliable estimate of the performance of the models was not implemented because of the imbalanced data distribution in the dataset. There are only 2 subjects who had very high BPRS scores in the dataset, so it was impossible to create multiple folds in a subject-independent manner that covered the whole spectrum of the variability of the BPRS scores in the dataset.

The aligned 40-second audio and video segments from the train split were used to train the MM-VQ-VAE-based MRL model as shown in Fig.\ref{fig:mrl}. The representation learning models are trained with a total loss which consists of audio reconstruction loss, video reconstruction loss, codebook loss (Commitment loss), and encoder loss \cite{Oord2017}. The model having the lowest total loss for the unseen test set was selected as the best-performing representation learning model. The best-performing representation learning model was trained with an embedding size of 1024, and a latent dimension size of 512 for 100 epochs with a learning rate of 1e-4. All these hyper-parameters were selected based on multiple experiments run with different combinations of the hyper-parameters.

To train the MM-VQ-VAE representation-based MTL downstream model as shown in Fig.\ref{fig:downstream}, the multimodal representations from the best-performing representation learning model for each 40-second segment were obtained. Then all the segment-wise representations belonging to a session were stacked together to form a 2D matrix which is used as the input for the downstream prediction model. To keep the input dimensions of the matrix consistent across sessions all the input matrices belonging to the shorter sessions were zero-padded to match the size of the longest session which created an input matrix size of $113\times1024$. To validate the need for the MTL paradigm, individual models were trained separately for the class prediction task with a Cross-Entropy loss and the severity score prediction task with MSE Loss. For the training of the combined MTL-based downstream prediction model, the total loss consisted of both the Cross-Entropy Loss and MSE Loss which were weighted using Homoscedastic Uncertainty Dynamic Loss Weighting as explained in the previous section of the paper. The downstream predication model was trained for 1000 epochs with a learning rate of 1e-5 and reduce by plateau learning rate scheduler with a threshold set on the validation loss of 0.001 with a patience of 150 epochs.

For the evaluation of the downstream predictions, Weighted F1(W.F1) score, AUC-ROC score, and Weighted Accuracy (W.ACC) were used for the 3-class classification, and Mean Absolute Error (MAE) was used for the severity prediction task.
\vspace*{-8pt}
\section{Results and Discussion}
\vspace*{-4pt}
The results of the downstream prediction tasks using our proposed methods and previous works done on the same dataset are shown in Table \ref{tab:performance}. The results show that when compared with multimodal models proposed in the previous studies, our proposed model of MM-VQ-VAE-based audio-visual speech representation worked better in the downstream tasks. This finding shows that properly trained multimodal representation learning models produce better representations than task-specific models with different multimodal fusion techniques. In addition, the results also indicate that the use of the MTL paradigm for training improves the performance of the classification task and the regression task.

\textbf{Error Analysis :}
Initially we investigated the performance of each session in the test-set on the severity estimation task. The average MAE score for the test set was mainly increased by sessions belonging to a single subject (SZ014). Subject SZ014's average BPRS score of 60 across all of its sessions was the highest across the whole dataset. And the average score was more than 2 standard deviations (2$\times$11.24) away from the mean BPRS score of the dataset (33.14). Just by removing this subject, we were able to obtain a 28.64\% drop in the MAE score as shown in Table \ref{tab:error}.
\vspace*{-8pt}
\begin{table}[h]
  \caption{\centering
  \textbf{Regression Error analysis}}
  \label{tab:error}
  \centering
  \begin{tabular}{ |l | c | }
  \hline
   \textbf{Test-set}&\textbf{MAE}\\
  \hline
    Original test-set with SZ014&7.19\\
    \hline
    Test-set without SZ014 and replaced by another SZ subject &5.59\\
    \hline
  \end{tabular}

\end{table}
\vspace*{-8pt}

Although the MM-VQ-VAE representation based MTL model has performed better than previous works, there is still room for improvement as the text modality was not included in this work. While the text modality was not the best-performing modality in our unimodal schizophrenia assessment frameworks introduced in previous studies \cite{premananth2024multimodal,premananth24_interspeech}, incorporating text in the multimodal frameworks has been proven to help increase the multimodal performance. In prior studies, the text modality was not aligned with other modalities to preserve the semantic structure of the text. However, aligning all input modalities is crucial for effective multimodal representation learning. Forcing the text to align with the segment lengths of audio and video would compromise its semantic integrity, potentially degrading the performance of schizophrenia-related tasks, as semantic coherence is known to be different between healthy controls and schizophrenia subjects \cite{https://doi.org/10.1002/wps.20491}. 
\vspace*{-2pt}
\section{Conclusion and Future Work}
\vspace*{-4pt}
In conclusion, we have demonstrated that the multimodal speech representations produced using TVs and FAUs from the MM-VQ-VAE representation learning framework are task-agnostic and work well on multiple downstream tasks. The use of a MTL paradigm combined with multimodal speech representations for the assessment of schizophrenia with severity prediction and symptom-based classification has proven to be beneficial in boosting performance across both tasks and have outperformed previous work. As of now, multimodal speech representation learning done for clinical use-cases is restricted to a single dataset because of limited data availability. However, in the future, we would like to incorporate data from different sources and datasets to produce more generalizable representations that can be used on different datasets and different tasks in the clinical domain. We are also planning to incorporate the text modality into the representation learning architecture to make the representations encompass more speech-related information.




\end{document}